\documentstyle[12pt]{article}

\begin{document}

\title{Gamma-Ray Bursts and Topology of the Universe}
\vspace{0.6 cm}
\author{
{\sc Marek Biesiada} \\
{\sl Copernicus Astronomical Center, } \\
{\sl Bartycka 18, 00-716 Warsaw, Poland}  \\
}

\date{September 1993}
\maketitle
\vfill
\begin{abstract}
\noindent

In this letter we propose a physical explanation for recently reported
correlations between pairs of close and antipodal gamma-ray bursts from
publicly available BATSE catalogue. Our model is based on the cosmological
scenario in which bursters are
located at cosmological distances of order of 0.5--2~Gpc. Observed
distribution of gamma-ray bursts strongly suports this assumption.
If so gamma-ray bursts may provide a very good probe for investigating the
topological structure of the Universe. We notice that correlation between
antipodal events may in fact indicate that we live in the so called Ellis'
small universe which has Friedman-Roberston-Walker metric structure and
nontrivial topology.

\end{abstract}
\vfill
\newpage

\section{Introduction}
Gamma ray bursts (GRBs thereafter) are undoubtly one of the most mysterious
phenomena in the sky with enormous diversity of durations, time variability and
spectra (for a review see \cite{PaczScience}).
Although their discovery was announced 20 years ago \cite{KSO}
and despite numerous efforts aimed at detecting these events (PVO, KONUS,
SIGNE, SMM, GINGA, BATSE) the nature of GRBs, mechanisms of their
$\gamma$-emission and the distance scales still remain unknown.
The last issue is crucial in a sense that determination of the distance scale
may in principle be performed without any detailed understanding of the GRBs
and will provide a severe constraint on the set of all possible models of GRBs
\cite{PaczPASCOS}.
Unfortunately all we can currently do is to make inferences about the distance
scales to the bursters from their distribution.\\
Since the work of Schmidt {\it et al.} \cite{Schmidt} GRBs' uniform
distribution is tested by extracting
a quantity $V \slash V_{max}$ which is a quotient of the volume determined by
the distance to the source and the maximal volume accessible for the detector.
For an isotropic and homogeneous population the distribution of $V \slash
V_{max}$ is uniform over a unit interval $[0,1]$ and has a mean value of
$< V \slash V_{max} > = 0.5$.

Thirteen years of continuous operating of PVO provided with a very good
statistic of strong bursts. They turned out to be distributed isotropically on
the celestial sphere (in angle) \cite{Atteia87} and uniformly in radial
coordinate ---
${< V \slash V_{max} >}_{PVO} = 0.46 \pm 0.02$ \cite{Fenimore}.
First results from the Burst and Transient
Source Experiment (BATSE) aimed at detection of weak bursters in hope to see
them concentrating towards the galactic plane were sensational \cite{Meegan}.
They revealed that the
distribution of weak GRBs is isotropic and that there is fewer weak bursts than
it could be expected from extrapolating the number of strong bursts ---
${< V \slash V_{max}>}_{BATSE} = 0.324 \pm 0.014$.
Hence we are apparently placed at the center of a spherically symmetric
distribution of GRBs which is uniform out to some distance and falls off
beyond.
This finding clearly prefers the hypothesis of cosmological origin of GRBs
\cite{cosmological}. Indeed we observe the Universe as isotropic and
the deficiency of weak (distant) bursts may be explained in a natural way
as a consequence of the Hubble expansion.
Although alternative explanations placing GRBs within the extended halo
\cite{halo} or the
Oort cloud \cite{oort} cannot be excluded definitely at this moment the
cosmological (extragalactic) scenario seems to be the most plausible one.

Recently Quashnock and Lamb \cite{QL} analysed the angular distribution of GRBs
from the BATSE catalogue using a nearest neighbor analysis. They found that
bursts are significantly clustered on an angular scale $\sim 4^o$.
Since systematic measurement errors in the BATSE experiment are of order of
$4^o$ they conjectured that GRBs typically repeat. If they were correct it
would rule out most of current extragalactic models which invoke single violent
events such like neutron star -- neutron star or neutron star -- black hole
mergers \cite{violent}.
Narayan and Piran \cite{NP} subsequently reanalysed the data using angular
autocorrelation function and repeated the nearest neighbor analysis of
Quashnock \& Lamb. They have employed the full sample of 260 bursts from BATSE
catalogue and defined a subsample (131 bursts) by including only those bursts
which have formal positional error smaller than $4^o$. The reason for the
latter choice is that in addition to the systematic error of $4^o$ bursts
positions have variable error which is estimated in the BATSE catalogue for
each event separately as the so called formal positional error (and which may
be as large as $20^o$). In addition a sample denoted in \cite{QL} as Type I+II
was considered in its full extent (201 events) and truncated (108 bursts) by
demand that the formal positional error be smaller than $4^o$.
What they found was an excess of close pairs with separations smaller than
$4^o$ as well as an excess of antipodal pairs with separations larger than
$176^o$. The statistical significance of the excesses was not impressive.
For example peaks of the autocorrelation function had amplitudes of
$1.75~\sigma$ for close pairs and $1.86~\sigma$ for antipodal pairs in the full
BATSE sample, $0.84~\sigma$ and $2.6~\sigma$ for truncated BATSE sample,
$1.95~\sigma$ and $1.75~\sigma$ for Quashnock \& Lamb sample and finally
$0.74~\sigma$ and $2.6~\sigma$ for truncated Quashnock \& Lamb sample.
Narayan and Piran concluded that both excesses (of close pairs and atipodal
ones) are likely due some unknown selection effect.

In the present letter we propose an explanation of reported excess of antipodal
GRBs provided that bursters are of cosmological origin and the Universe
posesses nontrivial topology.

\section{Topological structure of the Universe}

It is commonly accepted that the Universe we live in is extremely well
approximated by one of homogeneous and isotropic Friedman-Robertson-Walker
(FRW) models \cite{KolbTurner}.
The FRW metric may be written in the form:
\begin{equation} \label{FRWmetric}
ds^2 = c^2dt^2  - a^2(t)\bigl(\frac{dr^2}{1-kr^2} + r^2 d\theta^2 +
r^2\sin^2{\theta} d\varphi^2\bigr),
\end{equation}
where $a(t)$ is the scale factor and $k=0,+1,-1$ is the curvature of constant
time hypersurfaces and determines whether the Universe is flat, closed or open
respectively.
It is obvious that the metric of the space-time is of local nature and gives
no information about its topology. In other words a given metric structure such
like FRW metric (\ref{FRWmetric}) can be realized for different topologies.
For sake of illustration let us recall that two-dimensional flat Euclidean
metric can equally well be realized on a plane $R^2$, a cylinder or on a torus
$T^2.$ On the other hand all sucessfull physical predictions of standard
big-bang cosmology are based on the local metric structure (\ref{FRWmetric})
and the problem of topological structure of the Universe remains open.
All we know is that the Universe we observe is locally isotropic and
homogeneous i.e. the hypersurfaces of constant time $\Sigma_t$ are
3-dimensional spaces of constant curvature $k$ ($k=0,+1,-1$). Classification of
all topologically distinct spaces $\Sigma_t$ comes from the celebrated theorem
of Killing and Hopf that $\Sigma_t$ are isometric to $\tilde{\Sigma_t} \slash
\Gamma$ where $\Gamma$ is a discrete isometry subgroup of $\tilde{\Sigma_t}$
and $\tilde{\Sigma_t}$ (the so called covering manifold) is
\begin{quotation}
\noindent
$\tilde{\Sigma_t} = S^3$ --- three-sphere for k=+1,\\
$\tilde{\Sigma_t} = R^3$ --- Euclidean space for k=0,\\
$\tilde{\Sigma_t} = H^3$ --- three-dimensional hyperbolic space for k=-1.
\end{quotation}
The full topological classification (equivalent to enumerating all relevant
discrete groups $\Gamma$) exists for $k=0,+1$ \cite{Wolf} --- for $k=-1$ only
compact spaces can be classified \cite{Thurston}. Although the number of
topologically distinct spaces with FRW metric is large (18 for flat and
infinite for open and closed models) it is a common procedure to assume for
simplicity that $\Gamma = I$ and hence $\Sigma_t = \tilde{\Sigma_t}$.
Whenever $\Gamma$ is not equal to identity, the points equivalent under action
of $\Gamma$ are identified. Such a procedure generates a multiply connected
space in which geodesics connecting two points are not unique. This fact opens
the possibility of observationally verifying the topology of the Universe
provided the scale for multiply connectedness is smaller than present
horizon scale $a_0$ (the so called Ellis' small universe) \cite{Ellis} where
$a_0=c \slash H_0 \approx 3 \times 10^3 h_{100}^{-1} {\rm Mpc}$.

An unusual and distinctive feature of a multiply connected small
universe is that an observer sees many copies of the same object in different
directions at different distances.  The crucial point here is the
existence of geodesics starting and ending at the same point ---
the so called main geodesics of this point \cite{Demianski}.
Just to be specific let us restrict our attention to flat FRW universe ($k=0$)
with the topology of a torus $T^3$ generated by a discrete group $\Gamma$ of
traslations by vectors ${\bf e}_1,{\bf e}_2,{\bf e}_3.$
Then geodesics in the directions
${\bf d}_1 = \frac{{\bf e}_1}{|{\bf e}_1|},$
${\bf d}_2 =\frac{{\bf e}_2}{|{\bf e}_2|},$
${\bf d}_3 = \frac{{\bf e}_3}{|{\bf e}_3|},$ or
some linear combination of them ${\bf d}_{ij}=n_i{\bf d}_i + n_j
{\bf d}_j$ are the main geodesics.
If an object lies on (or near to) the main geodesic of an observer then two
copies of it may be observed in exactly (or nearly) opposite directions.
Another important effect is the periodicity of distances determined by the
radial coordinate $r$ in the FRW metric (\ref{FRWmetric}). Namely  if the
object is situated close to the main geodesic then the observer sees many
copies of it at distances $r+n|{\bf e}_i|$ where $n$ is an integer in
the direction ${\bf d}_i$ as well as in opposite direction.
More detailed discussion of the issue of main geodesics and observations in the
small universe can be found in \cite{Ellis} and \cite{Demianski}.

\section{Discussion and summary}
In this section we shall disucss an intriguing possibility that the effect of
correlation between antipodal pairs of GRBs reported by Narayan \& Piran
\cite{NP} may be an evidence for nontrivial topology of the Universe.
Let us start with known constraints on the scale $L_{top}$ for multiple
connectedness of the Universe. The earliest attempts to confront the idea of
multiply connected universe with observations performed by Sokolv \& Shvartsman
\cite{SS} and by  Gott \cite{Gott} constrained this scale from below ---
$L_{top}$ should be larger than $200~h_{100}^{-1}~ {\rm Mpc}$ where as usually
$h_{100}$ stands for the present Hubble constant in units of 100~km/s/Mpc and
is currently believed to lie between 0.4 and 1. This treshold value stems from
the fact that we do not observe many copies of familiar objects such like Coma
or Virgo clusters. Searching for periodicity in quasar distances Fang et al.
\cite{Fang} found positive effect
with periodicity scale (which is of the same order as $L_{top}$) $\sim 600~{\rm
Mpc}$. However they analysed only two small areas in the sky without
considering antipodal areas. Demia{\'n}ski and {\L}apucha \cite{Demianski}
investigated the effect of antipodal pairs of galaxies, clusters and quasars
and found marginally significant effect but gave no new estimate for $L_{top}$.
Therefore we shall adopt the estimate of Sokolov, Shvartsman \& Gott and
conclude that the effect of non-trivial topology of the Universe may
potentially manifest
itself only for objects at distances larger than $\sim 200~h_{100}^{-1}~{\rm
Mpc}$. This requirement is met for the population of GRBs in the cosmological
scenario.  The observed isotropy of GRBs combined with their uniform number
density out to some distance falling off beyond indicate that GRBs are most
likely at distances larger than $\sim 500~h_{100}^{-1}~{\rm Mpc}$. Indeed as
estimated by Mao \& Paczy{\'n}ski \cite{Mao} cosmological bursters should have
redshifts within the range from $z \approx 0.2$ to $z \approx 1.7$ which
correspond to the distance scale  of $\sim 500~h_{100}^{-1}~{\rm Mpc}$ and
$\sim 2~h_{100}^{-1}~{\rm Gpc}$ respectively (the second estimate comes from
the faintest bursts seen in BATSE). Hence the GRBs are very good candidates for
probing the topological structure of the Universe.

Although the statistical significance of excess in antipodal pairs of GRBs
found in \cite{NP} is not very impressive there is still some slight evidence
that this effect may be real. Whereas the significance of close pairs excess
depends crucially on the sample chosen the correlation of antipodal pairs
displays relative stability within tests performed in \cite{NP}. Moreover the
significance of excess in oposite pairs increases when the sample is improven
by rejecting bursts with large positional error unlike in the case of close
pair correlation. Narayan \& Piran did not commented on this since they assumed
that correlation of antipodal bursts is unphysical.

As already mentioned not every object observed in the small universe will have
its antipodal ghost image --- this effect manifests only in the direction of a
main geodesic. Therefore we may expect that only a certain fraction of GRBs
contributes to the net effect.
In order to be observed in the BATSE experiment the light coming from an GRB
directly and from the antipodes must arrive at approximately the same time.
In the other words two arcs of the main geodesic passing through an observer
and an GRB must have approximately the same length. This further constrains the
fraction of events contributing to antipodal pair correlation. If we knew the
topology of the universe in advance we might be able in principle to estimate
this fraction. However we may expect that very strong correlation
of only a fraction of events should leave an imprint on overall correlation
function even though the significance is decreased by contamination of
statistical sample by uncorrelated GRBs.

Narayan \& Piran noticed that
statistical significance would be by far better in the case of combined effect
of both close and oposite pairs correlation. Such an effect can easily be
explained in terms of periodic distance images close to main geodesics in
multiply connected universe. Indeed when the object ($\gamma$-burster) is
located near the main geodesic then we should see it acompanied by its ghost
images distributed periodically in radial comoving coordinate $r$ and the whole
picture should be reproduced at an antipodal locus. In the case of GRBs
it would mean in particular that strong bursts should be correlated with faint
ones. In fact such a correlation has been reported by Quashnock \& Lamb
\cite{QL}. However their nearest neighbor analysis was insensitive to antipodal
correlations. On the other hand Narayan \& Piran who discovered antipodal
correlations did not investigated correlations between faint and strong bursts.
Therefore future analysis of BATSE data in which correlation between close
and antipodal as well as between faint and strong bursts is carefully
investigated is a natural test for checking the idea proposed in this letter.
We can make a crude estimate of how many faint bursts may be expected to
accompany a strong event in the BATSE data. Since the strongest GRBs seen by
BATSE lie probably at $\sim 500~Mpc$ and the detector at Compton GRO is probing
the distance out to $\sim 2~Gpc$ then assuming the topological periodicty scale
$\sim 300~Mpc$ (which is a compromise between the lower bound of
Sokolov,Shvartsman \& Gott and the value reported by Fang) we may conclude that
up to about 5 weak events may be correlated with a strong one. This result is
however a very rough one depending crucially on the value of unknown scale
$L_{top}.$ Note also that topological periodicity scales can be different for
different directions.

It is obvious that in this scenario the faint burst must have occured about
$(L_{top} \slash 1\; Mpc) \times 3 \cdot 10^6\; yr \sim 10^9\; yr$
earlier than correlated strong one. Hence they cannot come from the same event.
On the other hand according to a merger scenario which is the most popular of
extragalactic scenarios GRBs are consequences of neutron star -- neutron star
or black hole -- neutron star mergers. As demonstrated in \cite{violent}
such a scenario is capable of explaining most of observed features of GRBs.
Suppose that a burst occured in certain galaxy then because the rate of mergers
is about $10^{-5} --10^{-6}\;yr^{-1}$ per galaxy we have a chance of accidental
coincidence of a given burst with one which have occured $\sim 10^9\;yr$
earlier and is now visible in one of ghost images of this galaxy.

In this letter we have proposed an explanation for reported recently
correlation between pairs of antipodal GRBs in the BATSE data. Our model
invokes a fascinating idea that our Universe may possess nontrivial topological
structure known for long as the so called Ellis' small universe.
Unfortunately angular resolution $\sim 4^o$ of the BATSE experiment is very
poor for verifying the hypothesis of small universe. It is very possible that
in the nearest future other
space experiments will provide accurate positions, of order of a fraction of an
arc minute, for strong bursts \cite{PaczScience}. This will enable us to test
some specific models of GRBs in the cosmological scenario. It will also be of
interest to see whether the oposite pairs of GRBs turn to be correlated within
this improved accuracy. More extensive statistical analysis of the BATSE data
will also tell us a lot.
It may well be the case that the original point of view of Narayan
\& Piran to search an explanation in some selection effect
is valid. Even if our hypothesis is disproven it is worth in our opinion to
recall from time to time and to reflect upon the fact that even though
we know the metric structure of the world we live in yet we can hardly say
anything about its topology. \\

When this letter has been completed a preprint by Maoz \cite{Maoz} appeared in
which author proposed an alternative explanation of antipodal pair correlation
--- the so called ring bias.
According to this hypothesis some bursts collapse to $4^o$ wide rings around
great circles
in the celestial sphere  because of GRO's localization procedure.
Bursts lying in the intersection of such rings would account for observed
correlation in both close and antipodal pairs.
It is worth noticing in the context of the present letter that
numerical simulations of Maoz clearly demonstrate that correlation between
close and antipodal pairs can be reproduced even if biased distribution is
mixed with randomly distributed GRBs. This supports our claim that it is
sufficient to have
only a fraction of events located suitably at main geodesics in the Ellis'
small universe in order to obtain observed features in autocorrelation
function.

\section*{Acknowledgements}
This project was supported by the Foundation for Polish Science and is a
contribution to the KBN Grant 2 20447 91 01.

\end{document}